\documentclass{Interspeech}

\interspeechcameraready

\title{%
    Towards Frame-level Quality Predictions of Synthetic Speech
}

\author[affiliation={1}]{Michael}{Kuhlmann}
\author[affiliation={2}]{Fritz}{Seebauer}
\author[affiliation={2}]{Petra}{Wagner}
\author[affiliation={1}]{Reinhold}{Haeb-Umbach}

\affiliation{}{Paderborn University}{Germany}
\affiliation{}{Bielefeld University}{Germany}
\email{%
    \{kuhlmann,haeb\}@nt.upb.de, \{fritz.seebauer,petra.wagner\}@uni-bielefeld.de
}

\usepackage{comment}
\usepackage{svg}
\usepackage{multirow}
\usepackage{tipa}
\usepackage{subcaption}
\usepackage{cleveref}
\usepackage[acronym]{glossaries}
\usepackage{cite}
\usepackage{xspace}
\usepackage{xurl}

\usepackage{amsmath}
\usepackage{amssymb}
\usepackage{bbm}
\usepackage{physics}
\usepackage{siunitx}
\usepackage{trfsigns}

\newcommand{\vect}[1]{\ensuremath{\boldsymbol{\mathbf{#1}}}}

\newcommand{\transpose}[1]{\ensuremath{#1^{\mathrm{T}}}}

\newacronym{dtw}{DTW}{dynamic time warping}
\newacronym{mos}{MOS}{mean opinion score}
\newacronym{pcc}{PCC}{Pearson correlation coefficient}
\newacronym{psds}{PSDS}{polyphonic sound detection score}
\newacronym{sed}{SED}{sound event detection}
\newacronym{sota}{SOTA}{state of the art}
\newacronym{srcc}{SRCC}{spearman rank correlation coefficient}
\newacronym{ssl}{SSL}{self-supervised learned}
\newacronym{ssqa}{SSQA}{subjective speech quality assessment}
\newacronym{tap}{TAP}{time average pooling}
\newacronym{wsola}{WSOLA}{waveform-synchronous overlap-add}

\newcommand\baselineA{SSL-MOS~\cite{Cooper2021Generalization}}
\newcommand\baselineAnc{SSL-MOS}
\newcommand\baselineB{SHEET SSL-MOS~\cite{Huang2024MOSBenchBG}}
\newcommand\baselineBnc{SHEET SSL-MOS}
\newcommand\reimpl{SHEET SSL-MOS: Re}
\newcommand\proposed{ChunkMOS}

\begin{document}

\maketitle

\begin{abstract}

    While automatic subjective speech quality assessment has witnessed much progress, an open question is whether an automatic quality assessment at frame resolution is possible. This would be highly desirable, as it adds explainability to the assessment of speech synthesis systems.  Here, we take first steps towards this goal by identifying issues of existing quality predictors that prevent sensible frame-level prediction. Further, we define criteria that a frame-level predictor should fulfill. 
    We also suggest a chunk-based processing that avoids the impact of a localized distortion on the score of neighboring frames. Finally, we measure in experiments with localized artificial distortions the localization performance of a set of frame-level quality predictors and show that they can outperform detection performance of human annotations obtained from a crowd-sourced perception experiment.
    

\end{abstract}

\section{Introduction}

The automatic \gls{ssqa} is an important task to track progress in the field of speech synthesis and to cut costs and time by being independent of crowd-sourced evaluation.
With the introduction of \gls{ssl} encoders~\cite{Cooper2021Generalization}, much progress has been made with the results well correlated with human perception~\cite{Huang2022TheVC, Huang2024TheVC, Tjandra2025AES}, and the focus now shifts to out-of-domain generalization~\cite{Huang2024TheVC, Huang2024MOSBenchBG}.
Both automatic and perceptual assessments summarize the audio quality of a query signal with a single scalar value, typically in the interval $[1,5]$, where $1$ denotes poor  and $5$  excellent quality~\cite{ITU85}.

To train such systems, utterances produced by speech synthesis systems are annotated with quality scores gathered from crowd-sourced perception experiments.
Although systems trained this way can successfully rank the overall quality of the query signal
\cite{Cooper2021Generalization,Huang2024TheVC, Tjandra2025AES}, current approaches do not look into explaining which parts of the signal contribute how much to the overall score.
For the further improvement of synthesis systems, we believe that it becomes increasingly important to assess synthesis quality at a more fine-grained level.
A hurdle in doing so is that there are no large-scale data sets for speech quality assessment with strong, i.e., frame-level, annotated quality scores.

So far, this problem has only been investigated by Quality-Net~\cite{Fu2018QualityNet}.
They used a BLSTM network with negative forget gate bias initialization to discourage the use of global context information across frames and a frame-level objective to regularize frame-level scores.
Although their analysis shares some similarities with ours, the forget gate bias initialization is not applicable to Transformer-based \gls{ssl} encoders which we are going to study.

In this work, we approach the problem of inferring strong from weak predictions as follows.
We start by identifying design choices of \gls{sota} \gls{ssqa} models that prevent us from extracting frame-level scores and show that there currently exists only one model from which frame-level scores are extractable.
This brings us to the definition of criteria that \emph{well-behaved} frame-level scores have to fulfill. We then introduce \proposed, a new \gls{ssqa} model to predict such well-behaved frame-level scores.
To test whether the above-identified and the proposed frame-level score prediction network are able to localize distortions, we inject synthetic, but realistic, distortions into clean utterances and  measure detection performance with an established metric from the sound event detection literature.
The source code is publicly available\footnote{\url{https://github.com/fgnt/frame-level-mos}}.

\section{%
    Predicting well-behaved frame-level scores
    from utterance-level targets
}

Neural \gls{ssqa} models follow an encoder/decoder paradigm~\cite{Wang2023RAMP,Huang2024MOSBenchBG}, where the encoder takes a raw speech signal as input and computes features with frame granularity, typically $\SI{50}{frames/s}$, which are further processed by the usually nonlinear decoder to yield an utterance-level scalar prediction on the MOS scale.
Common to all \gls{sota} predictors is the use of \gls{ssl} encoders such as wav2vec2~\cite{Baevski2020wav2vec2} or WavLM~\cite{Chen2021WavLM} for feature extraction.
We denote the sequence of hidden embeddings in the $l$-th layer, $l=1,\dots,L$, of such encoders by $\mathcal{H}^{(l)}=\{\vect{h}^{(l)}_1, \dots, \vect{h}^{(l)}_T\}$.
We start by analyzing the frame-level scores of \gls{sota} \gls{ssqa} models and show why they are unsuited for judging frame-level quality. 

\subsection{Analyzing frame-level scores of \gls{sota} \gls{ssqa} models}

The authors of~\cite{Cooper2021Generalization} were the first to introduce the use of \gls{ssl} encoders for \gls{ssqa}.
Their system performed equally well as more complicated systems in the VoiceMOS challenge 2022~\cite{Huang2022TheVC}.
All successful follow-up works relied on this \gls{ssl} approach~\cite{Huang2024TheVC, Tjandra2025AES} and recently, MOS-Bench~\cite{Huang2024MOSBenchBG}, a new benchmark to compare \gls{ssqa} systems, with the accompanying toolkit SHEET\footnote{\url{https://github.com/unilight/sheet}} was proposed, which offers pretrained models that build upon \baselineA.
Due to the success and popularity of \baselineAnc, we take it as the starting point for our analysis.

\subsubsection{Issue 1a: Unbounded frame-level scores}

\baselineAnc{} applies \gls{tap} and a linear projection to the outputs $\mathcal{H}^{(L)}$ of the last layer of a finetuned wav2vec2 encoder.
By rearranging the linear terms, we can write the predicted utterance-level score $o$ as
\begin{equation}
    o
    = \frac1{T}\sum_{t=1}^T \transpose{\vect{w}}\vect{h}^{(L)}_t +b = \frac1{T}\sum_{t=1}^Ts_t,
\end{equation}
where we introduced the frame-level score $s_t$.
Here, $\vect{w}\in\mathbb{R}^D$ is the learnable projection vector, $D$ is the dimension of the hidden embedding $\vect{h}_t^{(L)}\in\mathbb{R}^D$ at time frame $t$, and $b$ is a learnable scalar bias.
Neither utterance- nor frame-level scores are bounded and, in theory, can take values in the interval $[-\infty,\infty]$.
This is problematic as it will give different minimum and maximum scores across different utterances, preventing meaningful interpretation of frame-level scores.



The SHEET implementation of \baselineAnc~\cite{Huang2024MOSBenchBG} fixes this issue by (i) restricting the range of the scalar frame-level outputs with a $\tanh$ activation and applying proper scaling and shifting (range clipping) and (ii) performing \gls{tap} at the very last step of the model pipeline.
This gives the following modified notion for the frame-level scores of \baselineBnc:
\begin{equation}
    \label{eq:sheet-ssl-mos}
    s_t = \gamma\tanh\left(\transpose{\vect{w}}g\left(\vect{h}_t^{(L)};\theta_{\mathrm{dec}}\right)+b\right)+\beta,
\end{equation}
where $g(\cdot;\theta_{\mathrm{dec}}):\mathbb{R}^D\rightarrow\mathbb{R}^{D_g}$ is a nonlinear decoder network and $\gamma$ and $\beta$ are proper scaling and shifting scalars (e.g., ${\gamma=2}$ and $\beta=3$).
Note that the idea to perform range clipping was already proposed in the implementation of UTMOS~\cite{Saeki20222UTMOS}.
However, they proposed to apply the loss to each frame instead of the time-pooled utterance-level score.
We observed that this produced almost constant frame-level scores, which proved to be useless for judging frame-level quality.


\subsubsection{Issue 1b: Inaccessible frame-level scores}
Several works~\cite{Wang2023RAMP,Baba2024UTMOSv2, Tjandra2025AES} followed the original \baselineA{} by performing first \gls{tap}, but then processing the resulting embedding with a nonlinear decoder, giving the utterance-level score
\begin{equation}
    o = \transpose{\vect{w}}g\left(\frac1{T}\sum_{t=1}^T\vect{h}_t^{(L)};\theta_{\mathrm{dec}}\right)+b.
\end{equation}
Frame-level scores cannot be derived from such a model because $g(\cdot)$ and the \gls{tap} operation cannot be swapped
since the decoder is nonlinear here.
Hence, frame-level scores become \emph{inaccessible}.

\subsubsection{Issue 2: Local-global coupling behavior}
\label{sec:local-global-coupling}
\begin{figure}[t]
    \centering
    \begin{subfigure}[h]{\columnwidth}
        \centering
        \includegraphics[width=\columnwidth]{%
            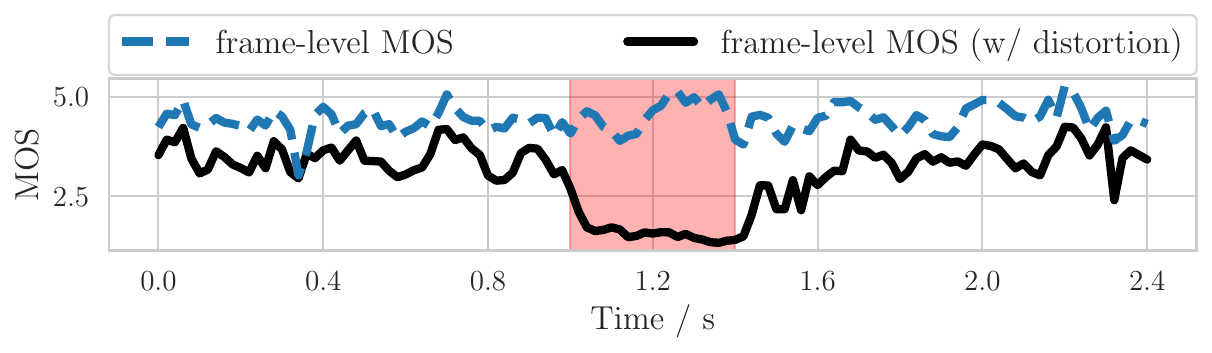
        }
    \end{subfigure}
    ~
    \begin{subfigure}[h]{\columnwidth}
        \centering
        \includegraphics[width=\columnwidth]{%
            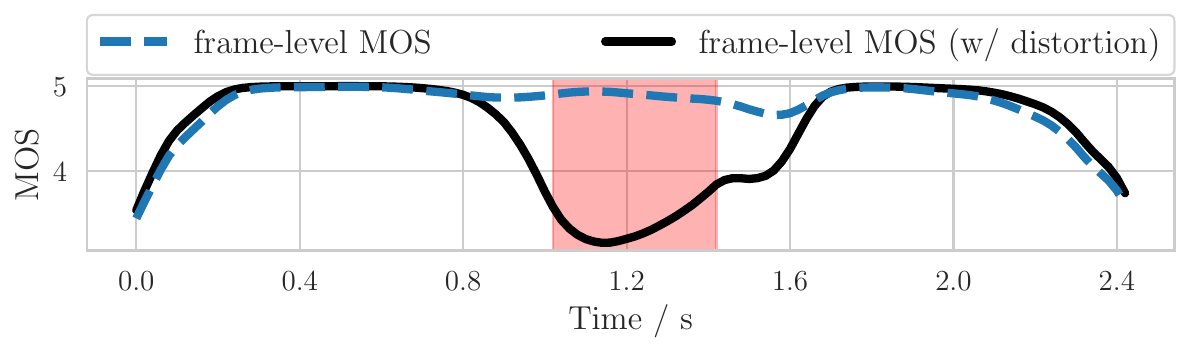
        }
    \end{subfigure}
    \caption{%
        Effect of adding a local distortion (pink noise) on frame-level MOS (blue: before, black: after adding the distortion).
        Because of the infinite receptive field of the Transformer, a local distortion additionally causes a mean shift (top), while it remains localized for \proposed\xspace (bottom).
    }
    \label{fig:local-global-coupling}
\end{figure}

While the above analysis focused on the decoder side, we also observed a crucial problem on the encoder side.
Consider a model with accessible frame-level scores such as the original \baselineAnc{} and a query signal containing a local distortion such as additive noise.
\Cref{fig:local-global-coupling}~(top) shows the frame-level output of \baselineAnc{} before and after applying the local distortion to the query signal.
Due to the infinite receptive field of the Transformer in the \gls{ssl} encoder, the local distortion causes not only a frame-level score drop in the degraded area but also a mean shift of the frame-level scores~\cite{Fu2018QualityNet}.
While this is not relevant from an utterance-level perspective, as long as the results correlate well with human perception, it becomes problematic when one is interested in attributing quality scores to frames:
A single local event can have unpredictable effects on independent frames.
We refer to this behavior of \gls{ssl}-based \gls{ssqa} models as \emph{local-global coupling}.

\subsection{On well-behaved frame-level scores}
Based on the above analysis, we now state conditions to ensure that frame-level scores are accessible and better interpretable when training with an utterance-level loss.
We define frame-level scores $\mathcal{S}=\{s_1,s_2,\dots,s_T\}$ obtained under weak target training as \emph{well-behaved} if they satisfy three conditions:
\begin{enumerate}
    \item \emph{Range clipping}: The range of frame-level scores is restricted to the MOS scale. 
    \item \emph{Accessibility}: Utterance-level scores are obtained by applying affine transformations to the pooled frame-level scores.
    \item \emph{Local-global decoupling}: Given two disjoint intervals $A$ and $B$ over $\mathcal{S}$, $A$ is invariant to changes in $B$. 
\end{enumerate}
Conditions 1 and 2 can be satisfied by following the design of \baselineBnc.
Since MOS annotated data are scarce, there is currently no other way than relying on \gls{ssl} encoders to obtain \gls{sota} performance.
Therefore, to satisfy condition 3, we propose to handle the input to the \gls{ssl} encoder differently.

\subsubsection{Decoupling the frame and utterance level}
We propose to chunk a query signal $\mathcal{X}=\{x_1,\dots,x_N\}$ into fixed-size blocks $\mathcal{B}=\{\mathcal{X}_1,\dots,\mathcal{X}_{\lceil\frac{N-B}{M}+1\rceil}\}$, each of length $B$ with a block shift of $0<M\leq{}B$ samples.
If a query signal is too short ($N<B$), we zero-pad it to length $B$.
Each chunk is processed independently by the \gls{ssl} encoder which can be efficiently done in a batch-wise manner.
If $M<B$, i.e., the blocks overlap, we perform overlap-add at the output of the \gls{ssl} encoder before forwarding the result to the decoder.

Additionally, instead of using a single block length, we propose to chunk the query signal at different resolutions $B_1,\dots,B_K$.
Denote the output of the \gls{ssl} encoder that receives inputs of block length $B_k$ by $\mathcal{H}_k=\{\vect{h}_{k1},\dots,\vect{h}_{kT}\}$ after applying overlap-add.
Then, the input to the decoder is a weighted sum of all outputs:
\begin{equation}
    \vect{h}_t = \sum_{k=1}^K w_k\vect{h}_{kt}, \quad \sum_{k=1}^Kw_k=1,
\end{equation}
where $w_1,\dots,w_K$ are learnable parameters of the model.
Note that this is conceptually similar to the spectrogram feature extractor in UTMOSv2~\cite{Baba2024UTMOSv2}, where a weighted sum over the outputs of multiple CNNs that received spectrograms at different time-frequency resolutions was taken.


\section{Experiments}
\label{sec:experiments}

\subsection{Tested frame-level predictors}
\label{sec:predictors}
We experiment with the following systems:
(a) as baselines, we take the pretrained checkpoints from the original \baselineAnc{}\footnote{\url{https://github.com/nii-yamagishilab/mos-finetune-ssl/blob/main/run_inference.py}} and its SHEET implementation\footnote{\url{https://huggingface.co/unilight/sheet-models/tree/main/bvcc/sslmos/2337}}, from which we extract frame-level scores;
(b) a reimplementation of \baselineBnc{} referred to as \reimpl, and (c) the reimplementation with chunked inputs, called \proposed.
We train \proposed{} with three block lengths of $\SI{1}{s}$, $\SI{600}{ms}$ and $\SI{400}{ms}$ with a shift of half a block length.
\reimpl{} differs from the SHEET recipe in the following points:
(i) we do not finetune the CNN feature extractor of the wav2vec2 encoder; (ii) we take a weighted sum of all hidden embeddings of the encoder~\cite{Baba2024UTMOSv2, Tjandra2025AES}, and (iii)
the utterance-level loss function is a combination of a clipped MSE loss~\cite{Leng2021MBNETMP} with clipping $\tau=0.1$ and a contrastive loss~\cite{Saeki20222UTMOS} with margin $\alpha=0.1$.
We found that these adaptations could improve the results.
As decoder, we  experiment with CNN and BLSTM architectures.
For the CNN, we use three layers with kernel size three, hidden size 512 and leaky ReLU activation.
For the BLSTM, we use three layers with hidden size 256 per direction.
We normalize the loudness of each utterance to $\SI{-18}{dBFS}$ before forwarding it to the \gls{ssl} encoder.
We train models (b) and (c) for $60$ epochs with an initial learning rate of $1\times10^{-5}$, which linearly decays to $1\times10^{-6}$ at the end of the training.
All models, including the baselines, were trained on the BVCC corpus~\cite{Cooper2021How}.
For evaluation, we take the checkpoint with the highest system-level \gls{srcc} on the validation data.

\subsection{Utterance- and system-level performance}
We start by comparing the systems on the utterance and system levels, following the evaluation protocol of the recent VoiceMOS challenges~\cite{Huang2022TheVC, Huang2024TheVC}.
Since out-of-domain performance is an important aspect of \gls{ssqa} models, we also evaluate on the test clean split of SOMOS~\cite{Maniati2022SOMOS}.
\Cref{tab:utterance-level} shows the results.
For the in-domain scenario, all models perform equally well, with \proposed{} performing a bit worse than \baselineAnc.
This gap becomes more extreme for the out-of-domain scenario, where \proposed{} clearly underperforms.

\begin{table}[t]
    \centering
    \caption{%
        Utterance- and system-level results of various MOS predictors for in-domain~(top) and out-of-domain~(bottom) data.
        Best results are given in {\bf{}bold}.
        Best results for each model family (resp. baselines) are \underline{underlined}. MSE: mean squared error; LCC: linear correlation coefficient.
    }
    \label{tab:utterance-level}
    \begin{subtable}[h]{\columnwidth}
    \label{tab:utterance-level-bvcc}
    {
        \footnotesize
        \setlength{\tabcolsep}{3.5pt}
        \begin{tabular}{lcccccc}
            \toprule[1.5pt]
             & \multicolumn{3}{c}{\bf{}Utterance-level} & \multicolumn{3}{c}{\bf{}System-level}  \\
             \cmidrule(lr){2-4}\cmidrule(lr){5-7}
             Model & MSE & LCC & SRCC & MSE & LCC & SRCC \\
             \midrule[1pt]
             \baselineA & .277 & \underline{.869} & .869 & .145 & .924 & .922 \\
             \baselineB & \underline{.271} & .867 & \underline{.870} & \underline{.123} & \underline{.930} & \underline{.931} \\
             \midrule
             \reimpl & {\bf\underline{.230}} & .870 & .872 & {\bf\underline{.088}} & {\bf\underline{.933}} & {\bf\underline{.930}} \\
             \quad + CNN & .248 & {\bf\underline{.879}} & {\bf\underline{.878}} & .116 & .922 & .921 \\ 
             \quad + BLSTM & .303 & .869 & .867 & .147 & .918 & .911 \\ 
             \midrule
             \proposed & .537 & .851 & .856 & .320 & .916 & .914 \\
             \quad + CNN & .550 & \underline{.858} & \underline{.859} & .322 & \underline{.922} & \underline{.916} \\
             \quad + BLSTM & \underline{.490} & .855 & .855 & \underline{.275} & .915 & .911 \\
             \bottomrule[1.5pt]
        \end{tabular}
    }
    \end{subtable}
    \begin{subtable}[h]{\columnwidth}
    \vspace{1em}
    \label{tab:utterance-level-somos}
    {
        \footnotesize
        \setlength{\tabcolsep}{3.3pt}
        \begin{tabular}{lrccrcc}
            \toprule[1.5pt]
             & \multicolumn{3}{c}{\bf{}Utterance-level} & \multicolumn{3}{c}{\bf{}System-level}  \\
             \cmidrule(lr){2-4}\cmidrule(lr){5-7}
             Model & MSE & LCC & SRCC & MSE & LCC & SRCC \\
             \midrule[1pt]
             \baselineA & \underline{.640} & {\bf\underline{.481}} & {\bf\underline{.490}} & \underline{.453} & \underline{.662} & {\bf\underline{.708}} \\
             \baselineB & {{.849}} & {.451} & {.467} & {.614} & {.643} & {.691} \\
             \midrule
             \reimpl & {\bf\underline{.546}} & \underline{.476} & \underline{.457} & {\bf\underline{.356}} & {\bf\underline{.683}} & \underline{.693} \\
             \quad + CNN & .601 & .467 & .450 & .416 & .681 & \underline{.693} \\ 
             \quad + BLSTM & .716 & .458 & .445 & .537 & .641 & .667 \\ 
             \midrule
             \proposed & 1.296 & .348 & .346 & 1.125 & .512 & .582 \\
             \quad + CNN & 1.345 & \underline{.374} & \underline{.375} & 1.174 & .537 & \underline{.612} \\
             \quad + BLSTM & \underline{1.272} & .367 & .363 & \underline{1.088} & \underline{.541} & .601 \\
             \bottomrule[1.5pt]
        \end{tabular}
    }
    \end{subtable}
\end{table}

\subsection{Local-global coupling behaviors}
\label{sec:experiments.local-global-coupling}
In~\Cref{sec:local-global-coupling}, we have reasoned the local-global coupling behavior to be problematic for interpreting frame-level scores.
Here, we are performing a large-scale quantitative evaluation to show that this problem indeed occurs.
We take utterances from the test-clean split of LibriTTS-R~\cite{Koizumi2023LibriTTS-R}, add a $\SI{1}{s}$ long pink noise distortion to each utterance and compute the frame-level scores.
Around the distortion, we apply a collar of $\SI{200}{ms}$ and take the left and right subsequences of the frame-level scores.
We extract these subsequences before and after applying the distortion.
To compare the similarity of the subsequences before and after distortion, we calculate their \gls{pcc} and their distance given by the minimum alignment cost from \gls{dtw}~(\Cref{tab:local-global}).

Clearly, the chunked processing effectively resolves the local-global coupling, in particular when using CNN or BLSTM decoder layers. Not only is the \gls{pcc} the highest, the low \gls{dtw} costs indicate that frame-level score trajectories are very similar without and with a distorted segment in the neighborhood.
This is also exemplified for a single utterance in~\Cref{fig:local-global-coupling}~(bottom).
For \baselineBnc{}, we made observations similar to those for UTMOS~\cite{Saeki20222UTMOS}, i.e., the frame-level scores are mostly constant, leading to a more prominent mean shift.
This has extreme detrimental effects on the detection performance (see below).
\begin{table}[t]
    \centering
    \caption{%
        Change of frame-level scores after introducing a distortion. 
        We measure Pearson correlation coefficient and dynamic time warping alignment cost of the unaffected subsequences (i.e., to the left and to the right of the distortion) before and after applying the distortion.
        Best results are given in {\bf{}bold}.
    }
    \label{tab:local-global}
    {
        \footnotesize
        \setlength{\tabcolsep}{7.5pt}
        \begin{tabular}{lccrr}
            \toprule[1.5pt]
            Model & lPCC & rPCC & lDTW & rDTW \\
            \midrule[1pt]
            \baselineA & .792 & .789 & 32.99 & 28.48 \\
            \baselineB & .377 & .386 & 144.61 & 114.70 \\
            \midrule
            \reimpl & .804 & .772 & 66.81 & 80.63 \\
            \quad + CNN & .880 & .842 & 38.43 & 43.58 \\
            \quad + BLSTM & .681 & .659 & 67.84 & 46.04 \\
            \midrule
            \proposed & .867 & .822 & 1.82 & 2.88 \\
            \quad + CNN & {\bf.924} & {\bf.905} & 2.42 & {\bf3.31} \\
            \quad + BLSTM & .913 & .809 & {\bf1.73} & 3.89 \\
            \bottomrule[1.5pt]
        \end{tabular}
    }
\end{table}

\subsection{Frame-level performance}
There exist no dataset with frame-level quality predictions which can serve as ground truth for frame-level prediction. For this reason, we added synthetic localized distortions to a clean speech signal and evaluated how well the frame-level predictor can localize the corrupted frames.
We acknowledge that this approach has deficiencies because the introduced distortions are artificial.
Nevertheless, this serves as an indication whether frame-level predictions from utterance-level annotated training data are, in principle, possible.

\subsubsection{Performance measure}
To measure the localization accuracy of the detected events, we employ metrics from the evaluation of sound event detection systems.
Specifically, we use the intersection-based criterion~\cite{Bilen2019Framework} to classify a prediction as a hit, miss, or false alarm, depending on how large the intersection of the detected segment containing a distortion is with the ground truth: If the intersection is larger than $\rho_{\textrm{DTC}}$, where $\rho_{\textrm{DTC}}\in[0,1]$, it is considered a hit, else a miss, and a false alarm is indicated if there was no distorted frame in the ground truth annotation of the detected region.
To be independent of the detection threshold, we report the area under the receiver operating characteristic curve (AUC).

\subsubsection{Artificial distortions}
\label{sec:artificial_distoritons}

{\noindent\bf{}Additive noise.}
We use additive noise to simulate synthesis artifacts in the generation of fricatives.
As such, distortion areas are sampled where the forced alignment indicates a {fricative}
and we add {\bf{}pink noise} with standard deviation $\sigma=0.1$.

{\noindent\bf{}Phase artifacts.}
We compute the STFT of a single distortion area and replace the phase vectors with {\bf{}random phase} vectors drawn from the uniform distribution $\mathcal{U}(-\pi,\pi)$.
For phase randomziation, we constrain distortion areas to voiced segments.

{\noindent\bf{}Prosody artifacts.}
We simulate two types of prosody artifacts.
First, we {\bf{}shift the pitch} by either a quarter or half octave.
We constrain distortion areas for pitch shifting to segments where a pitch was detected.
Second, we {\bf{}stretch the duration of vowels} by a random factor from the uniform distribution $\mathcal{U}(2,3)$.

The injection is done as follows:
For each utterance to be evaluated, we randomly sample a distortion class.
Next, we sample $N=3$ distortion areas such that the onset of an area aligns with the constraints of the sampled distortion and apply this distortion to all areas.
We sample the duration for each area from the uniform distribution ${\mathcal{U}(\SI{400}{ms}, \SI{700}{ms})}$ which corresponds to the average duration of an English word stem.

\subsubsection{Detection performance}
\label{sec:experiments.detection}

To measure detection performance, we use the intersection-based criterion with ${\rho_{\textrm{DTC}} = 0.5}$
and ${\rho_{\textrm{DTC}}= 0.7}$,
where the latter penalizes less accurate detections more severely.
Following post-processing in sound event detection, we apply median filters of varying duration ($\SI{0}{ms}$ to $\SI{500}{ms}$ in steps of $\SI{50}{ms}$) and choose the filter length with the best AUC on the dev-clean split.
We use \texttt{sed\_scores\_eval}\footnote{\url{https://github.com/fgnt/sed_scores_eval}}~\cite{Ebbers2022Threshold} for hyperparameter tuning and AUC metric computation.
\Cref{tab:detection} shows the results.

\setlength{\tabcolsep}{12pt}
\begin{table}[t]
    \centering
    \caption{%
        Intersection-based detection results of artificially distorted clean speech utterances from LibriTTS-R~\cite{Koizumi2023LibriTTS-R} test-clean, averaged over all distortion classes.
        Best results are highlighted in {\bf{}bold}.
        Best results for each model family (resp. baselines) are \underline{underlined}.
    }
    \label{tab:detection}
    {
        \footnotesize
        \setlength{\tabcolsep}{8pt}
        \begin{tabular}{@{\extracolsep{\fill}}lcccc}
            \toprule[1.5pt]
            Model & \multicolumn{2}{c}{{$\rho_\textrm{DTC}=0.5$}} & \multicolumn{2}{c}{$\rho_\textrm{DTC}=0.7$} \\
            \cmidrule(lr){2-3}\cmidrule(lr){4-5}
            & medfilt & AUC & medfilt & AUC \\
            \midrule[1pt]
            \baselineA & \SI{400}{ms} & \underline{.650} & \SI{350}{ms} & \underline{.513} \\
            \baselineB & \SI{450}{ms} & .028 & \SI{200}{ms} & .007 \\
            \midrule
            \reimpl & \SI{300}{ms} & .603 & \SI{250}{ms} & .451 \\
            \quad + CNN & \SI{450}{ms} & {\bf\underline{.683}} & \SI{300}{ms} & {\bf\underline{.551}} \\
            \quad + BLSTM & \SI{450}{ms} & .595 & \SI{350}{ms} & .451 \\
            \midrule
            \proposed & \SI{500}{ms} & .544 & \SI{500}{ms} & .266 \\
            \quad + CNN & \SI{500}{ms} & .622 & \SI{500}{ms} & .317 \\
            \quad + BLSTM & \SI{500}{ms} & \underline{.682} & \SI{500}{ms} & \underline{.341} \\
            \bottomrule[1.5pt]
        \end{tabular}
    }
\end{table}

The results indicate that detection of distortions is possible using weakly obtained frame-level scores.
For ${\rho_{\textrm{DTC}}=0.5}$,
we achieve competitive performance with chunked inputs and a BLSTM decoder compared to an encoder with a global attention window.
When detection requirements become stricter (${\rho_{\textrm{DTC}}=0.7}$),
\proposed{}
falls behind by a large margin.
We made two observations that may explain this inexact localization.
First, frame-level scores of \proposed{} are inert, especially in combination with a BLSTM decoder, tending to predict too long activity, as can be seen in the comparison in~\Cref{fig:local-global-coupling}.
Second, drops in the frame-level score curve tend to lag such that a local minimum falls together with the offset of a distortion, decreasing the possible overlap to a maximum of $50\%$.



\subsection{%
    Comparison to human detection performance
}
In a final experiment, we compare the distortion localization to that of human annotations.
In a small-scale listening experiment similar to~\cite{Seebauer2023re}, we asked 40 laypeople to mark areas they thought were detrimental to the overall quality of the speech signal.
In total, 18 distorted utterances (6 per distortion class) from the EARS~\cite{Richter2024EARS} corpus were queried.
We counted how many participants marked a frame as detrimental and normalized this count by the number of ratings for that utterance to obtain soft scores.
To tune the hyperparameters, we performed a 5-fold cross validation using the \texttt{SEBBs} toolbox\footnote{\url{https://github.com/merlresearch/sebbs}}~\cite{Ebbers2024Sound}.
\Cref{tab:comp-subjective} shows the results.
We find that weakly obtained frame-level scores from all models outperform the localization performance of laypeople, especially for strict detection requirements (${\rho_{\textrm{DTC}}=0.7}$).
This outcome does not encourage us to gather crowd-sourced frame-level targets but instead to further pursue weak learning of frame-level scores.
\begin{table}[t]
    \centering
    \caption{%
        Detection performance of human annotations versus frame-level scores of \gls{ssqa} models on 18 distorted utterances from the EARS~\cite{Richter2024EARS} corpus.
    }
    \label{tab:comp-subjective}
    {
        \footnotesize
        \setlength{\tabcolsep}{7pt}
        \begin{tabular}{lcccc}
			\toprule[1.5pt]
			Method & \multicolumn{2}{c}{$\rho_{\textrm{DTC}}=0.5$} & \multicolumn{2}{c}{$\rho_{\textrm{DTC}}=0.7$} \\
			\cmidrule(lr){2-3} \cmidrule(lr){4-5}
			& F1 & AUC & F1 & AUC \\
			\midrule[1pt]
			Human Annotated & .630 & .361 & .274 & .064 \\
			\midrule
			  \baselineA & .656 & .454 & .359 & .234 \\
			\midrule
			\reimpl+CNN & {\bf.672} & {\bf.543} & {\bf.376} & {\bf.332} \\
			\proposed+BLSTM & .586 & .408 & .288 &  .193 \\
			\bottomrule[1.5pt]
		\end{tabular}
    }
\end{table}

\section{Discussion and Outlook}
We have shown that weak learning of frame-level scores for \gls{ssl}-based \gls{ssqa} is possible and might be the best way to obtain these given the discouraging results of the crowd-sourced perception experiment.
Although \proposed{} is the only model that predicts \emph{well-behaved} frame scores, it cannot yet achieve the detection performance of models that produce ill-behaved frame-level scores.
Improving the performance of \proposed{} is of ongoing interest.
Ultimately, our goal is to apply the idea of frame-level scores to actual synthetic speech to discover system-specific error patterns.

\section{Acknowledgements}
\ifinterspeechfinal
    This research was funded by Deutsche Forschungsgemeinschaft (DFG), project 446378607.
    Computational resources were provided by the Paderborn Center for Parallel Computing.
\else
    Acknowledgements hidden during review.
\fi

\bibliographystyle{IEEEtran}
\bibliography{mybib}

\end{document}